\documentclass[aps,prl,reprint,amsmath,amssymb,superscriptaddress]{revtex4-2}
\usepackage{bibunits}
\defaultbibliographystyle{apsrev4-2}
\defaultbibliography{HQC_bib}
\usepackage{graphicx}
\usepackage{amsmath,amssymb}
\usepackage{braket}
\usepackage{bm}
\usepackage{xcolor}
\usepackage{verbatim}
\usepackage{enumitem}
\usepackage{multirow}
\usepackage[colorlinks=true,allcolors=blue]{hyperref}
\graphicspath{{./}{upload/}}
\usepackage{physics}
\usepackage{amsmath,amssymb}

\newcommand{\fourstar}{4-Star}
\newcommand{\sixring}{6-Ring}

\begin{document}

\begin{bibunit}
    
\title{High-Threshold Fusion-Based Quantum Computation with Optical Hybrid Qubits}

\author{Seong-Yeop Lim}
\affiliation{Department of Physics and Astronomy, Seoul National University, Seoul 08826, Korea}
\author{Jaehak Lee}
\affiliation{Center for Quantum Technology, Korea Institute of Science and Technology (KIST), Seoul 02792, Korea}
\affiliation{Division of Quantum Information, KIST School, Korea University of Science and Technology (UST), Seoul 02792, Republic of Korea}
\author{Nuri Kang}
\affiliation{Department of Applied Physics, Yale University, New Haven, Connecticut 06511, USA}
\author{Seung-Woo Lee}
\email{swleego@gmail.com}
\affiliation{Department of Physics, Pohang University of Science and Technology (POSTECH), Pohang 37673, Korea}

\begin{abstract}
\noindent 
Hybrid discrete- and continuous-variable (DV–CV) encodings have been developed to overcome operational drawbacks of each individual encoding, while fusion-based quantum computation (FBQC) has emerged to address the practical challenges of scalability toward fault-tolerant photonic quantum computation. Here we introduce hybrid FBQC using small optical DV–CV hybrid resource states connected through hybrid fusions, to combine the advantages of both approaches. We develop protocols for directly generating these resource states without requiring local hybrid logic such as the Hadamard gate, together with a tailored decoding strategy that exploits the error information contained in individual hybrid-fusion outcomes. We demonstrate that hybrid FBQC achieves a photon loss threshold of 2.231\% while requiring only 12 hybrid fusions per surface-code unit cell, substantially improving upon the 0.89\% threshold and 60-fusion requirement of previous hybrid measurement-based scheme. These results establish optical hybrid FBQC as a high-threshold resource-efficient route to scalable fault-tolerant photonic quantum computation.
\end{abstract}

\maketitle


Photonic quantum computing has developed along either discrete-variable (DV) or continuous-variable (CV) routes ~\cite{Kok2007,OBrien2009,Slussarenko2019}. DV encodings with single photons provide orthogonal logical bases and accurate qubit manipulation, but rely on probabilistic entangling gates and Bell measurements that increase resource overhead. In contrast, CV encodings support efficient Gaussian operations and measurements, but suffer from non-orthogonal bases and difficulty generating non-Gaussian states. These complementary features have motivated hybrid approaches that combine DV and CV advantages to jointly encode or process quantum information~\cite{Loock2011,Andersen2015,LeeSW2013,Jaehak2024,Bose2024,Jaehak2026}.

Meanwhile, photonic platforms are particularly well suited to measurement-based quantum computation (MBQC) thanks to their modularity and fast measurements~\cite{OBrien2009,Raussendorf2001}.
Fault-tolerant MBQC architectures typically requires preparing a three-dimensional cluster state and performing local measurements to implement quantum error correction and logical operations~\cite{Raussendorf2006,Raussendorf2007,Dawson2006,Herrera2010,LeeSH2023}. 
Building on this framework, several photonic DV--CV hybrid MBQC protocols have been  developed~\cite{Omkar2020,Omkar2021,Jaehak2024}. Topological hybrid MBQC has been demonstrated using single photons and coherent states ~\cite{Omkar2020,Omkar2021}. Hybridizing a bosonic cat code with single photons for an outer topological code was proposed to improve both loss tolerance and resource overhead by combining the merits of CV and DV~\cite{Jaehak2024}.

A central challenge of MBQC, however, is the preparation of large entangled states, such as the Raussendorf-Harrington-Goyal (RHG) lattice implementing the surface code, which requires many probabilistic entangling operations and thus introduces substantial resource overhead and additional error locations~\cite{Dawson2006,Herrera2010,Bartolucci2023}. This difficulty is also pronounced in hybrid architectures~\cite{Omkar2020,Omkar2021,Jaehak2024}. In particular, even microcluster preparation requires additional hybrid logic, such as gate teleportation to implement Hadamard gates, which introduces further entangled resources and fusion operations~\cite{Jaehak2024}. 
Fusion-based quantum computation (FBQC) was recently proposed to circumvent this bottleneck of MBQC by replacing a preassembled universal cluster state with small entangled resource states connected through fusions  whose outcomes directly yield error-correction checks~\cite{Bartolucci2023}. Subsequent proposals have improved the fault tolerance of FBQC through optimized resource geometries, code constructions, and encoding-based fusion strategies~\cite{Bombin2023,Sahay2023,Paesani2023,Song2024}. However, a hybrid FBQC scheme combining the complementary advantages of DV and CV encodings while directly exploiting hybrid-fusion outcomes for quantum error correction has yet to be explored.

In this Letter, we fill this gap by introducing hybrid FBQC scheme using optical DV--CV hybrid qubits. By replacing large hybrid cluster states with only small-sized resource states, our scheme offers a realistic and feasible route to scalability. 
The modular structure of the fusion network further allows the DV and CV components to be implemented separately, offering a natural route to heterogeneous and distributed architectures. 
Beyond these architectural advance, we develop techniques tailored to optical hybrid FBQC. We devise efficient protocols for directly generating the requisite resource states without the teleportation-based local hybrid operations such as Hadamard gate~\cite{Jaehak2024}. This direct construction removes the additional fusion steps and associated pathways for error propagation. We further develop an outcome-aware decoding strategy that exploits the error information carried by individual hybrid-fusion outcomes.
As a result, our scheme substantially improves fault tolerance while reducing both resource costs and fusion overhead. Specifically, it achieves a loss threshold of 2.231\% while requiring only 12 hybrid fusions per surface-code unit cell, representing a substantial improvement over previous hybrid MBQC schemes, which achieve 0.89\% threshold with 60 fusions per unit cell~\cite{Jaehak2024}. Notably, these results surpass previous CV and hybrid architectures in both fault tolerance and resource efficiency~\cite{Jeong2002,Omkar2021,Larsen2021b,Bourassa2021,Tzitrin2021,Jaehak2024,Jaehak2026}.
The attained thresholds also numerically exceed those reported for the original single-photon FBQC~\cite{Bartolucci2023}, suggesting further improvements through strategies such as boosting~\cite{Bartolucci2023} and encoded-fusion~\cite{LeeSW2015,Paesani2023,Song2024} on the DV side. 
Our work establishes hybrid FBQC as a high-threshold resource-efficient route to scalable fault-tolerant quantum computation, and lays the groundwork for heterogeneous and distributed DV--CV quantum architectures.


\begin{figure}
    \centering
    \includegraphics[clip=true, width=0.95\linewidth]{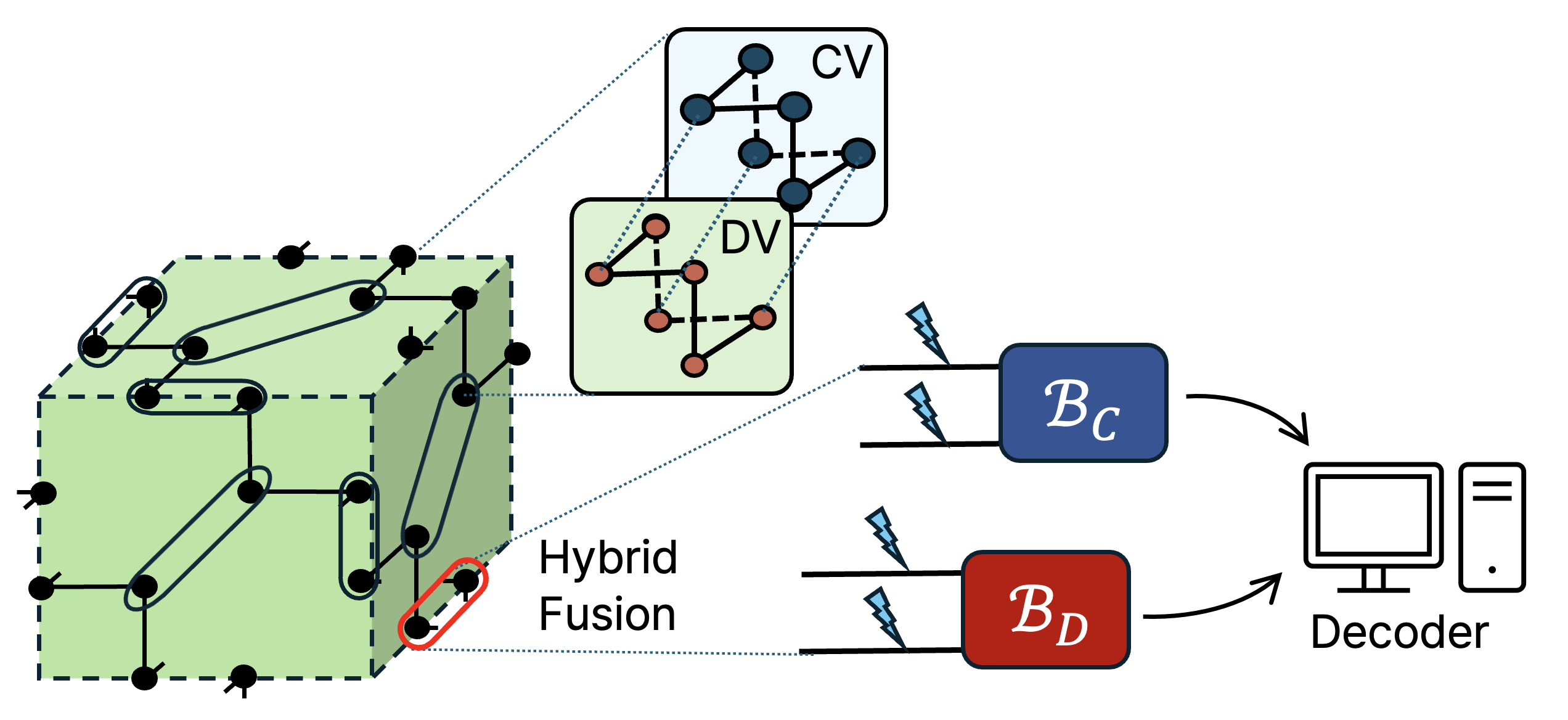}
    \caption{Optical hybrid FBQC. A unit cell of the 6-ring fusion network and its corresponding 6-ring resource state are shown as a representative architecture. Small hybrid resource states are connected via hybrid fusions, in which DV and CV Bell measurements are performed independently. The resulting measurement records are jointly processed by a decoder for quantum error correction.
    } 
    \label{fig:hybrid_arch}
\end{figure}

\textit{Optical hybrid encoding.--}
We first introduce an optical DV-CV hybrid qubit. In our approach, each hybrid qubit consists of a single-photon polarization state and a CV mode with the logical basis
\begin{equation}
\ket{0_L} = \ket{+}\ket{\phi_0},\qquad
\ket{1_L} = \ket{-}\ket{\phi_1},
\label{eq:hybrid_basis}
\end{equation}
where $ \ket{\pm} = (\ket{H} \pm \ket{V})/\sqrt{2} $. For CV component, we consider coherent state encoding, $(\ket{\phi_0},\ket{\phi_1})=(\ket{\alpha},\ket{-\alpha})$, denoted H-coh, and cat-code encoding, $(\ket{\phi_0},\ket{\phi_1})=(\ket{C_\alpha^+},\ket{C_{i\alpha}^+})$, denoted H-cat, where $ \ket{C_\alpha^\pm} \propto |\alpha\rangle \pm | -\alpha \rangle$. 
The DV component guarantees the orthogonal logical basis, overcoming the basis-overlap inherent in CV-only encoding~\cite{Cochrane1999, Jeong2002, Ralph2003, Leghtas2013, Mirrahimi2014, Bergmann2016, Li2017, Ourjoumtsev2007}. On the other hand, the CV component enhances Bell-state discrimination as $\alpha$ increases, allowing the fusion success probability to exceed that of conventional DV-only linear-optical fusion~\cite{LeeSW2013,Jaehak2024}. In addition, the cat-code in H-cat enables the detection of single-photon loss through a change in photon-number parity from the even parity $\ket{C_\alpha^+}$ to the odd $\ket{C_\alpha^-}$~\cite{Leghtas2013,Mirrahimi2014,Grimsmo2020}.

By construction, the hybrid Bell states can be decomposed into linear combinations of products of DV and CV Bell states as
\begin{equation}
\label{eq:hybridBS}
\begin{aligned}
    & \ket{\Phi_L^\pm} \propto \ket{\Phi_D^+}\ket{\tilde{\Phi}_C^\pm} + \ket{\Psi_D^+}\ket{\tilde{\Phi}_C^\mp},\\
    & \ket{\Psi_L^\pm} \propto \ket{\Phi_D^-}\ket{\tilde{\Psi}_C^\pm} + \ket{\Psi_D^-}\ket{\tilde{\Psi}_C^\mp},
\end{aligned}
\end{equation}
where $ \ket{\Phi_D^\pm} =(\ket{HH} \pm \ket{VV})/\sqrt{2}$ and $ \ket{\Psi_D^\pm}=(\ket{HV} \pm \ket{VH})/\sqrt{2}$  are the single-photon Bell states, while $\ket{\tilde{\Phi}_C^\pm} $  and $ \ket{\tilde{\Psi}_C^\pm}$ denote the corresponding unnormalized CV Bell states for the H-coh or H-cat encoding.

The DV and CV components can therefore be measured separately via Bell measurements $\mathcal{B}_D$ and $\mathcal{B}_C$, respectively, with their classical outcomes processed jointly to infer the logical observables $X_L\otimes X_L$ and $Z_L\otimes Z_L$, denoted by $XX$ and $ZZ$, respectively. The Bell-state letter, $ \Psi(\Phi) $, is assigned if $ ZZ = +1(-1) $ , while its sign $\pm$ is assigned if $ XX = \pm 1 $. Accordingly, combining the outcomes of $\mathcal{B}_D$ and $\mathcal{B}_C$ identifies the hybrid Bell state. 
$\mathcal{B}_D$ is implemented by a Type-II polarization Bell measurement~\cite{Browne2005, Lut99, Calsamiglia2001}, which can unambiguously distinguish only two of the four Bell states with linear optics. The complementary information provided by $\mathcal{B}_C$ can resolve this remaining ambiguities left by $\mathcal{B}_D$, thereby enhancing the overall hybrid-fusion success probability beyond that of the DV measurement alone. For $\mathcal{B}_C$, H-coh employs coherent-state Bell measurement, whereas H-cat employs cat-code Bell measurement. For H-cat fusion, we consider two protocols proposed by Hastrup and Andersen  (HA)~\cite{Hastrup2022} and Su, Dhand, and Ralph (SDR)~\cite{Su2022}, both based on linear optics and photon-number-resolving detection (PNRD) by observing measurement patterns. Details of these fusion schemes are provided in the Supplemental Material.
Beyond enhancing the fusion success probability, individual measurement records from $\mathcal{B}_D$ and $\mathcal{B}_C$ provide side information that can be exploited for quantum error correction in the hybrid fusion network, as discussed below.

\textit{Hybrid fusion architecture.--}
FBQC performs fault-tolerant quantum computation by connecting small, constant-size stabilizer states through fusion measurements whose outcomes collectively define the error-correction checks~\cite{Bartolucci2023, Bombin2023}. For 4-Star and 6-Ring resource states, repeated fusions generate a three-dimensional syndrome graph corresponding to the spacetime foliation of the surface code~\cite{Raussendorf2006, Raussendorf2007, Fowler2009, Bartolucci2023}. Its vertices represent parity checks, while the edges are associated with fusion outcomes, such that errors on the fusion outcomes are detected through violations of the neighboring checks. This structure naturally accommodates the optical hybrid qubits considered here. In hybrid FBQC, each fusion edge is implemented by a hybrid fusion consisting of $\mathcal{B}_D$ and $\mathcal{B}_C$ as illustrated in Fig.~\ref{fig:hybrid_arch}. Once the resource states are prepared, error correction proceeds entirely through fusion measurements, without requiring additional local hybrid operations. This removes, in particular, the hybrid Hadamard operations required in previous hybrid-MBQC constructions, which are nontrivial to implement for the bosonic cat-state component~\cite{Lund2008, Omkar2020, Omkar2021, Jaehak2024}. Moreover, unlike the hybrid-MBQC construction, fusion errors do not propagate through a preassembled cluster of surviving data qubits before decoding, but are directly associated with edges of the fusion syndrome graph.

\begin{figure}[t]
    \centering
    \includegraphics[clip=true, width=\linewidth]{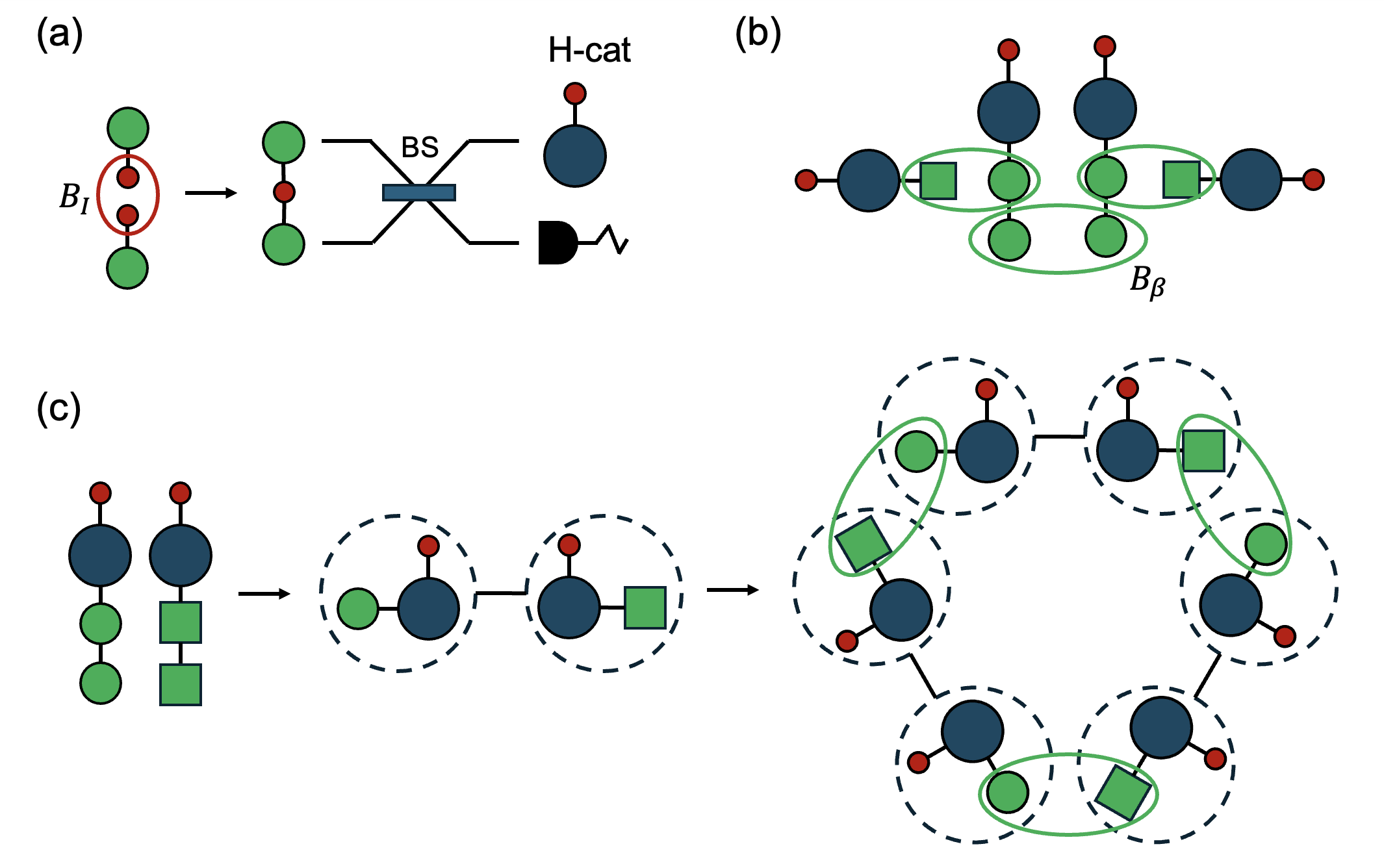}
    \caption{Direct H-cat resource preparation. Red, green, and blue circles denote single-photon polarization, coherent-state, and cat-code component, respectively. Green squares denote even/odd cat superpositions of ancillary coherent states. (a) H-cat preparation using Type-I fusion $\mathcal{B}_I$ and even-parity postselection. (b) A hybrid 4-Star state is generated by merging two $\ket{\phi_T^{(1)}}$ and two $\ket{\psi_T^{(2)}}$ with ancillary coherent Bell fusions $\mathcal{B}_\beta$. (c) A hybrid 6-Ring state is generated by megring three $\ket{\psi_T^{(2)}}$ and three $\ket{\phi_T^{(2)}}$. Detailed preparation schemes are given in the Supplemental Material.}
    \label{fig:resource}
\end{figure}

\textit{Direct resource-state generation.--}
We next devise schemes for directly preparing the hybrid resource states required for FBQC without local hybrid gate operations.
We take a H-coh state, $\ket{\mathrm{H\mbox{-}coh}}=(\ket{+}\ket{\alpha}+\ket{-}\ket{-\alpha})/\sqrt{2}$, as the elementary resource unit, which has been demonstrated experimentally~\cite{Jeong2014,Morin2014,Guccione2020,Darras2023,Ulanov2017,Sychev2018}.
H-coh resource states can be assembled using the hybrid-state preparation primitives developed for coherent-state hybrid computation~\cite{LeeSW2013,Omkar2020,Omkar2021} (see Supplemental Material).
We thus focus here on the preparation of H-cat resource states.
An H-cat state is prepared from two H-coh states by applying a Type-I fusion $\mathcal{B}_I$ to the DV components and a balanced beam splitter followed by PNRD to the CV components as illustrated in Fig.~\ref{fig:resource}(a). Since both operations succeeds with probability approximately $1/2$, the average cost of generating one H-cat state is approximately eight H-coh states~\cite{Hastrup2020, Jaehak2024}.

The target resource states can then be constructed from H-cat states augmented with $n$ ancillary coherent-state modes in two complementary forms,
\begin{align}
    |\psi_T^{(n)}\rangle \propto|0_L\rangle |\beta\rangle^{\otimes n} + |1_L\rangle |-\beta\rangle^{\otimes n}, \\
    |\phi_T^{(n)}\rangle \propto|0_L\rangle |\beta_+\rangle^{\otimes n} + |1_L\rangle |\beta_-\rangle^{\otimes n},
\end{align}
where $|\beta_\pm\rangle \propto |\beta\rangle \pm |-\beta\rangle$. 
Applying coherent-state Bell fusion $\mathcal{B}_\beta$ between ancillary modes from the two families implements a logical CZ operation up to a Pauli-frame correction determined by the fusion outcome~\cite{Jaehak2024}.
This enables the required entangled graph states to be constructed directly, without applying hybrid Hadamard gates.
The preparation schemes for the 4-Star and 6-Ring resource states are illustrated in Fig.~\ref{fig:resource}(b,c). 
The modified 4-Star state is constructed from two $\ket{\psi_T^{(2)}}$ and two $\ket{\phi_T^{(1)}}$ blocks using fusion in the $\{XX,ZZ\}$ basis. Note that the 4-Star state is here modified to have the stabilizer generators  $ \{ X_1X_2Z_3Z_4, \ Z_1Z_2, \ Z_2X_3, \ X_3X_4 \} $. Likewise, the 6-Ring state is constructed from three $\ket{\psi_T^{(2)}}$ and three $\ket{\phi_T^{(2)}}$ blocks. The resource costs can be estimated from the failure probability $1/2$ for $\mathcal{B}_I$ and $e^{-2|\beta|^2}$ for $\mathcal{B}_\beta$, the latter vanishes as $\beta$ increases. Further details of the resource-state preparation and cost analysis are provided in the Supplemental Material.

\begin{table}[t]
\centering
\caption{Decision rules and conditional error probabilities for hybrid-fusion measurements. Columns correspond to DV measurement success, failure, and photon-loss erasure. 
The decision row specifies the logical fusion outcomes inferred from the DV and CV measurement results,
while the remaining rows give the corresponding $ZZ$- and $XX$-error probabilities.}
\label{tab:hybrid_fusion_outcomes}
\renewcommand{\arraystretch}{1.35}
\setlength{\tabcolsep}{3pt}
\scriptsize
\begin{tabular}{@{}c|ccc@{}}
\hline\hline
\textbf{DV}
& \begin{tabular}{@{}c@{}}
    \textbf{Success}\\[-2pt]
    ($\Psi$)
  \end{tabular}
& \begin{tabular}{@{}c@{}}
    \textbf{Fail}\\[-2pt]
    ($\Phi$)
  \end{tabular}
& \begin{tabular}{@{}c@{}}
    \textbf{Erased}\\[-2pt]
    (photon loss)
  \end{tabular}
\\
\hline
\textbf{Decision}
& \begin{tabular}{@{}c@{}}
    $ZZ_L = XX_{\mathcal B_D}$\\
    $XX_L = -XX_{\mathcal B_C}$
  \end{tabular}
& \begin{tabular}{@{}c@{}}
    $ZZ_L = ZZ_{\mathcal B_C}$\\
    $XX_L = XX_{\mathcal B_C}$
  \end{tabular}
& \begin{tabular}{@{}c@{}}
    $ZZ_L = ZZ_{\mathcal B_C}$\\
    $XX_L = XX_{\mathcal B_C}$
  \end{tabular}
\\[2pt]
\hline
\textbf{$ZZ$ error}
& $0$
& \multicolumn{2}{c@{}}{%
    \begin{tabular}{@{}cl@{}}
      $p^{ZZ}$ & if ambiguous $ZZ$ in $\mathcal B_C$ \\
      $0$ & otherwise
    \end{tabular}%
  }
\\[2pt]
\hline
\textbf{$XX$ error}
& $p^{XX}$
& $p^{XX}$
& $\frac{1}{2}$
\\
\hline\hline
\end{tabular}
\end{table}

\textit{Hybrid fusion outcomes.--}
We now show how the individual records from $\mathcal{B}_D$ and $\mathcal{B}_C$ are used to infer hybrid fusion outcomes and extract outcome-dependent error probabilities.
Here we consider $\mathcal{B}_D$ that unambiguously distinguishes $\ket{\Psi_D^+}$ and $\ket{\Psi_D^-}$, but cannot distinguish $\ket{\Phi_D^+}$ from $\ket{\Phi_D^-}$. 
Without photon loss, the detection patterns always determine the DV $ZZ$ outcome, whereas the $XX$ outcome is resolved only for $\ket{\Psi_D^\pm}$. 
If either single photons is lost, the DV measurement information is erased. Accordingly, its outcomes are classified as success (identification of $\ket{\Psi_D^\pm}$), failure (unresolved $\ket{\Phi_D^\pm}$), or erasure (heralded photon loss), which can be distinguished from the detection patterns of $\mathcal{B}_D$.

For CV fusion $\mathcal{B}_C$, H-coh employs coherent-state Bell measurement, whereas H-cat uses the HA or SDR cat-fusion protocol. In H-cat fusion, the outcomes are inferred from PNRD patterns~\cite{Hastrup2022,Su2022}. However, certain patterns do not uniquely determine the $ZZ$ outcome, as they are compatible with both possible $ZZ$ values. In such cases, the measurement projects the input onto a superposition $\mu\ket{\Phi_C^\pm}+\nu\ket{\Psi_C^\pm}$. We assign the more probable Bell component as the inferred outcome, with a conditional $ZZ$-error probability $p^{ZZ}=\min{(|\mu|^2,|\nu|^2)}/(|\mu|^2+|\nu|^2)$. In addition, multiphoton loss can induce $XX$ flips, some of which can be identified from the detection patterns, while others remain unheralded with a probability $p^{XX}$~\cite{Jaehak2024}.
The cat-code qubit exhibits a biased-noise property: $ p^{ZZ} $ is suppressed exponentially with increasing amplitude $ \alpha $, whereas $ p^{XX} $ increases only linearly with $ |\alpha|^2 $.
Details of the fusion circuits, measurement patterns, and corresponding error probabilities are provided in the Supplemental Material.

The measurement records from $\mathcal{B}_D$ and $\mathcal{B}_C$ provide complementary information, allowing hybrid-fusion outcomes to be inferred even when one component measurement is inconclusive. The corresponding decision rules are summarized in Table~\ref{tab:hybrid_fusion_outcomes}.
When $\mathcal{B}_D$ succeds, the hybrid-fusion outcomes are inferred as $\{ZZ_{L}= XX_{\mathcal{B}_D}, \, XX_L=-XX_{\mathcal{B}_C} \}$. 
In this case, the DV measurement record allows a CV $ZZ$ flip or ambiguity to be resolved through the ideal relation $XX_{\mathcal{B}_D}=ZZ_{\mathcal{B}_C}$ from Eq.~(\ref{eq:hybridBS}), ensuring the correct hybrid $ZZ$ outcome is inferred.
For example, if $\mathcal{B}_D$ identifies $\ket{\Psi_D^+}$ and $\mathcal{B}_C$ yields $\ket{\Phi_C^+}$, the hybrid Bell state is identified as $\ket{\Phi_L^-}$ from Eq.~\eqref{eq:hybridBS}. Even if $\mathcal{B}_C$ yields $\ket{\Psi_C^+}$, the DV result reveals a $ZZ$ flip in CV, allowing the same hybrid Bell state $\ket{\Phi_L^-}$ to be correctly inferred.
When $\mathcal{B}_D$ fails or is erased, the hybrid-fusion outcomes are instead inferred from the CV record using $\{ ZZ_L = ZZ_{\mathcal{B}_C},\, XX_L= XX_{\mathcal{B}_C} \}$. 
In these cases, a $ZZ$ flip in $\mathcal{B}_C$ cannot be resolved using the DV measurement record, resulting in a non-zero $ZZ$ error probability, $p^{ZZ}$.
On the other hand, a CV $XX$ flip remains a hybrid $XX$ error regardless of whether $\mathcal{B}_D$ succeeds or fails, with a conditional error probability $p^{XX}$. Furthermore, when $\mathcal{B}_D$ is erased by photon loss, the hybrid $XX$ outcome becomes completely undetermined, giving $p^{XX}=1/2$.

\textit{Outcome-aware decoding.--}
Beyond determining the hybrid-fusion outcomes, the individual records from $\mathcal{B}_D$ and $\mathcal{B}_C$ provide information about residual fusion errors. We exploit this information in outcome-aware decoding by assigning measurement-conditioned error probabilities to the syndrome graph.
Under independent pure-loss channels with loss rate $\eta$, each hybrid fusion yields a record $m=(m_{D},m_{C})$, comprising the DV measurement status and the CV PNRD pattern. Since the complete loss events are not directly accessible, we assign the outcome-conditioned error probabilities
\begin{equation}
   p^E(m)= P(E \mid m;\eta,\alpha),\quad E\in\{XX,ZZ\},
\end{equation}
by averaging over unobserved loss sectors conditioned on the measurement record. For example, a successful $\mathcal{B}_D$ removes the $ZZ$ ambiguity, $p^{ZZ}(m)=0$, whereas DV photon loss erase the $XX$ outcome, i.e.,~$p^{XX}(m)=1/2$.

These probabilities are then mapped to the corresponding syndrome-graph edges, labeled by $e$. This mapping is direct for the 6-Ring network, while each effective edge in the 4-Star network is determined by the parity of four sub-fusion errors. The resulting edge-error probabilities $p_e^E$ determine the minimum-weight perfect matching (MWPM) weights,
\begin{equation}
  w_e^E=\log \frac{1-p_e^E}{p_e^E}.
\end{equation}
In contrast, outcome-unaware decoding discards the individual measurement records and uses the averaged fusion-error probabilities, $\bar p^E=\sum_m P(m | \eta,\alpha)P(E | m;\eta,\alpha)$. Details of the conditional error probabilities and their mapping to syndrome-graph weights are provided in the Supplemental Material.



\begin{figure}[t]
    \centering
    \includegraphics[clip=true, width=\linewidth]{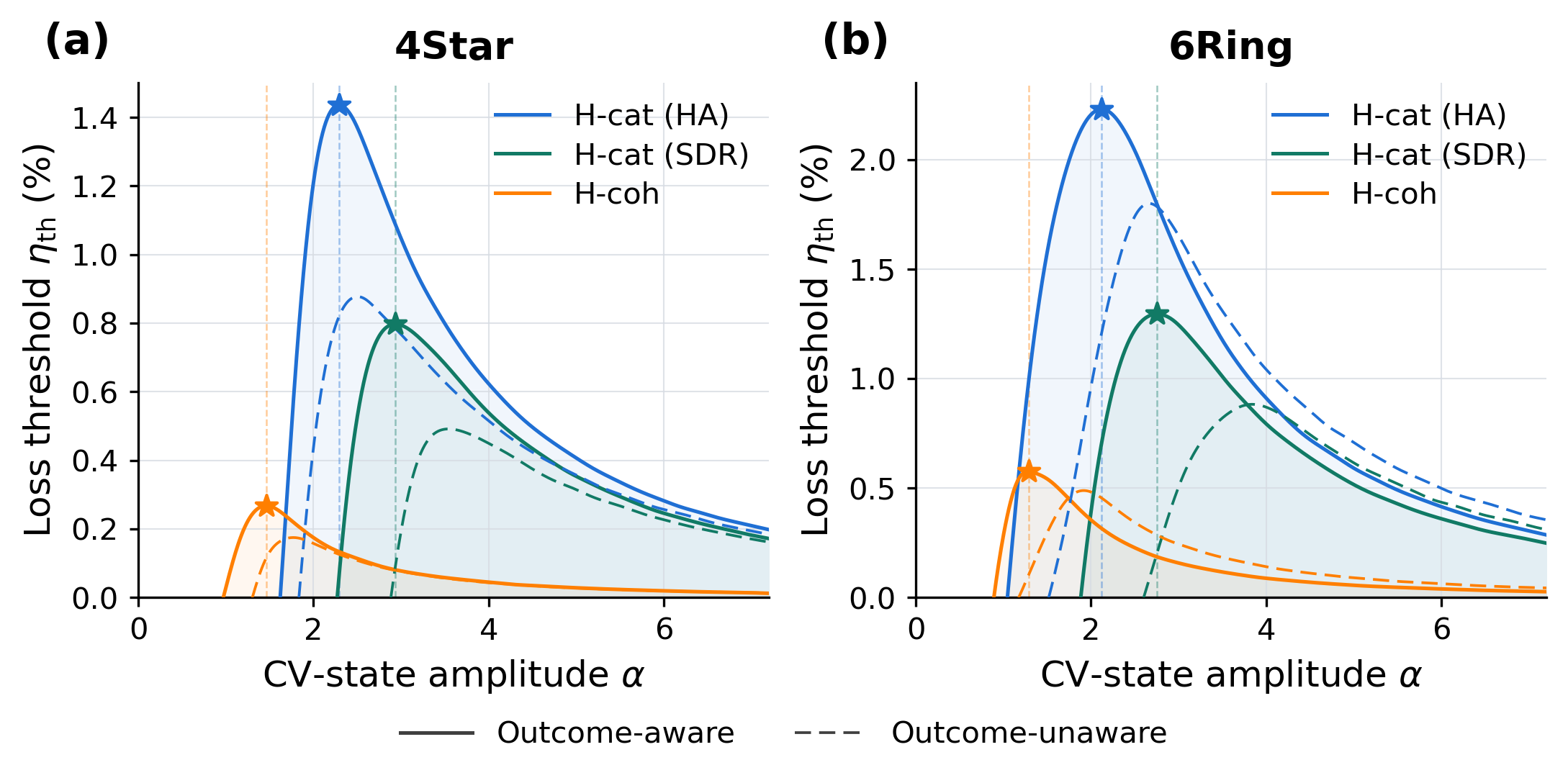}
    \caption{ Photon loss threshold $\eta$ versus CV-state amplitude $\alpha$ for the 4-Star and 6-Ring hybrid-FBQC architectures. Solid and dashed lines denote outcome-aware and outcome-unaware decoding respectively. Star marks the maximum threshold of each curve.}
    \label{fig:threshold}
\end{figure}

\textit{Fault-tolerance analyses.--}
We evaluate the loss tolerance of the 4-Star and 6-Ring hybrid-FBQC architectures using H-coh and H-cat encodings.
We assume independent loss channels with an identical loss rate $\eta$ for all optical modes and ideal resource-state preparation, while accounting for the intrinsic fusion failure and measurement ambiguity. 
We perform Monte Carlo simulations with MWPM decoding for code distances $d=7, 9, 11$. 
A logical error occurs when the residual error chain after decoding is topologically nontrivial.
The loss threshold is estimated from the crossings of logical-error-rate curves for different code distances~\cite{Fowler2012}.
For each parameter point, samples are accumulated until the half-width of the 90\% Clopper-Pearson confidence
interval is below 10\% of the estimated logical error rate.

In Fig.~\ref{fig:threshold}, we plot the loss thresholds against the CV-state amplitude $\alpha$, comparing outcome-aware
(solid lines) and outcome-unaware (dashed lines) decoding for both architectures.
The threshold exhibits a nonmonotonic dependence on $\alpha$. For small amplitude, the finite overlap of the CV basis states increases the ambiguity of the cat-fusion measurement, whereas increasing $\alpha$ improves CV Bell-state distinguishability and suppresses the associated fusion errors~\cite{Hastrup2022,Su2022,Jaehak2024}.
At large amplitude, however, increasing mean photon number makes the CV states more susceptible to photon loss, leading to an optimal intermediate amplitude for each architecture and fusion scheme.

Without outcome information, the 4-Star architecture achieves loss thresholds of 0.877\% (HA), 0.492\% (SDR), and 0.175\% (H-coh), while the corresponding 6-Ring thresholds are 1.800\%, 0.883\%, and 0.489\%.
Incorporating individual fusion-measurement records substantially improves the thresholds.
With outcome-aware decoding, the H-cat thresholds increase to 1.436\% (HA) and 0.797\% (SDR) for the 4-Star network, and to 2.231\% (HA) and 1.295\% (SDR) for the 6-Ring network. Compared with the corresponding outcome-aware H-coh
results, the HA thresholds are $5.35$ and $3.88$ times higher for the 4-Star and 6-Ring networks, respectively.
These results demonstrate the advantage of exploiting side information of the hybrid fusion outcomes into the classical decoder provides a decisive advantage in the hybrid FBQC scheme. We also show that the 4-Star and 6-Ring networks require only 24 and 12 hybrid fusions per surface-code RHG unit cell, respectively. Notably, the 6-Ring architecture achieves a loss threshold of 2.231\% with only 12 hybrid fusions per unit cell, compared with 0.89\% and 60 fusions in the previous hybrid-MBQC scheme~\cite{Jaehak2024}.


\textit{Discussion and outlook.--}
We have introduced optical hybrid FBQC that combines the advantages of DV--CV hybrid encoding and fusion-based architectures for fault-tolerant photonic quantum computation. By replacing the large hybrid cluster states with small hybrid resource states, our scheme offers a more experimentally feasible and efficient route to scalability. We have developed 
protocols for directly preparing these resource states without local hybrid operations, such as Hadamard gates, thereby eliminating additional fusion steps and associated pathways for error propagation.
We have also devised an outcome-aware decoding strategy that exploits the error information contained in individual DV and CV fusion outcomes. As a result, our hybrid FBQC substantially improves loss tolerance while reducing both resource costs and fusion overhead. Specifically, 4-Star and 6-Ring architectures achieve loss thresholds of 1.436\% and 2.231\%, respectively, with only 24 and 12 hybrid fusions per surface-code unit cell, compared with 0.89\%. and 60 fusions in previous MBQC~\cite{Jaehak2024}. These results also outperform previous CV and hybrid architectures in both fault tolerance and resource cost, based on the comparison in Ref.~\cite{Jaehak2024,Jaehak2026}.

Several directions may further improve the fault-tolerance of hybrid FBQC. Our results show that encoding the CV side in a bosonic cat code (H-cat) raises the threshold further than the coherent-state encoding. A natural next step would be investigating the performance with multicomponent cat codes~\cite{Bergmann2016, Li2017, Grimsmo2020}, provided that efficient hybrid-fusion protocols can be developed. Moreover, the attained thresholds numerically exceeds those of the original single-photon FBQC~\cite{Bartolucci2023}, although the resource cost are not directly comparable. This suggests that possibility of further improvements through DV-side fusion-enhancement strategies, such as boosting~\cite{Bartolucci2023} and encoded-fusion~\cite{Paesani2023,Song2024}. 
 
Beyond fault-tolerance improvement, the modular structure of hybrid FBQC offers opportunities for heterogeneous and distributed quantum architecture. The DV and CV components can, in principle, be implemented in distinct subsystems and connected through hybrid-fusion interfaces, potentially enabling distributed quantum computation accross physically separated platforms. Recent experimental progress in generating and manipulating optical DV--CV hybrid states provides promising building blocks for such implementations~\cite{Morin2014,Guccione2020,Darras2023}. We further expect that the extension need not be limited to all-optical platforms, but could involve other physical platforms such as ion traps, neutral atoms, and superconducting circuits~\cite{Awschalom2021,Hann2019,Gan2020,Young2022,Crane2024,Jaehak2026}.
Our work thus provides a efficient framework for combining hybrid encoding with FBQC and opens possibilities for scalable and distributed DV--CV quantum architectures.

This research was funded by National Research Foundation of Korea (RS-2022-NR068812, RS-2024-00442762), Institute of Information \& Communications Technology Planning \& Evaluation (RS-2025-02263264), Korea Institute of Science and Technology (2E33541,2E33546), and Global Partnership Program of Leading Universities in Quantum Science and Technology (RS-2025-02317602).

\putbib
\end{bibunit}


\clearpage
\begin{bibunit}
    
\onecolumngrid
\begin{center}
    \textbf{Supplementary Materials} 
\end{center}
\twocolumngrid

\setcounter{equation}{0}
\section{Optical fusion building blocks}
\label{sec:fusion_building_blocks}

Hybrid FBQC uses four optical measurement primitives. The hybrid fusion itself consists of a Type-II DV fusion $\mathcal{B}_D$ and a cat-code Bell measurement $\mathcal{B}_C$. Resource-state preparation additionally uses a Type-I DV fusion $\mathcal{B}_I$ and a coherent-state Bell measurement $\mathcal{B}_\beta$. In this section, we summarize these operations and the information contained in their measurement outcomes. Our conventions follow the hybrid-qubit construction and optical fusion analysis of Ref.~\cite{Jaehak2024}.

\subsection{DV fusion}
\label{sec:DV_fusion}

\begin{figure}[t]
    \centering
    \includegraphics[width=\linewidth]{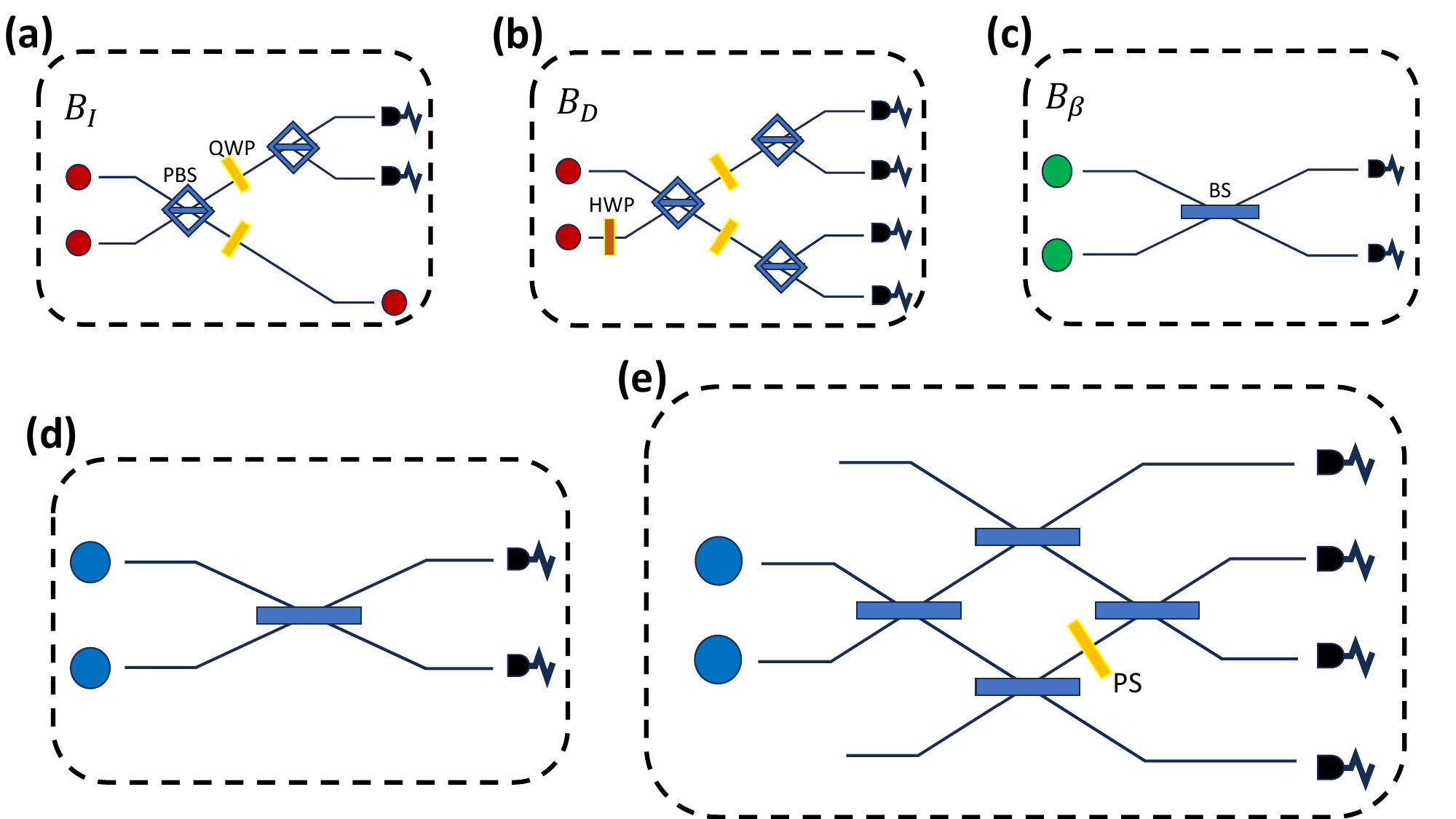}
    \caption{Optical fusion primitives. (a) Type-I DV fusion $\mathcal{B}_I$. PBS: polarizing beam splitter; QWP: quarter-wave plate. (b) Type-II DV fusion $\mathcal{B}_D$. HWP: half-wave plate. Photon-number resolution distinguishes loss from the intrinsic Bell-measurement failure. (c) Coherent-state Bell measurement $\mathcal{B}_\beta$ \cite{LeeSW2013}. BS: balanced beam splitter. (d) HA cat-code Bell measurement \cite{Hastrup2022}. (e) SDR cat-code Bell measurement \cite{Su2022}; PS denotes a $\pi/2$ phase shifter.}
    \label{fig:fusion_primitives}
\end{figure}

The DV component is encoded in the polarization of a single photon. Type-I and Type-II fusions are linear-optical partial Bell measurements \cite{Browne2005,Lut99,Calsamiglia2001}. The Type-I fusion $\mathcal{B}_I$, shown in Fig.~\ref{fig:fusion_primitives}(a), maps two polarization qubits to one output qubit when exactly one photon is detected. Up to a known Pauli-frame correction, its successful Kraus action can be written as
\begin{equation}
    \ket{+}\!\bra{HH}_{A,B}+\ket{-}\!\bra{VV}_{A,B},
\end{equation}
for the convention used in our resource-state-generation circuit. In the ideal lossless setting, $\mathcal{B}_I$ succeeds with probability $1/2$.

The Type-II fusion $\mathcal{B}_D$, illustrated in Fig.~\ref{fig:fusion_primitives}(b), performs an incomplete Bell measurement on two polarization qubits. We define
\begin{align}
    \ket{\Psi_D^\pm}&=\frac{\ket{H}\ket{V}\pm\ket{V}\ket{H}}{\sqrt{2}},\\
    \ket{\Phi_D^\pm}&=\frac{\ket{H}\ket{H}\pm\ket{V}\ket{V}}{\sqrt{2}}.
\end{align}
With linear optics, $\mathcal{B}_D$ distinguishes the two $\ket{\Psi_D^\pm}$ outcomes. When the detection pattern corresponds to the $\Phi_D$ subspace, the Bell letter is known but the sign is unresolved; we call this a heralded DV-fusion failure. In the ideal lossless case this occurs with probability $1/2$ \cite{Browne2005,Lut99,Calsamiglia2001,Jaehak2024}. If either input photon is lost, fewer than two photons are detected and the DV Bell information is completely erased. Photon-number-resolving detectors therefore distinguish a successful outcome, a heralded fusion failure, and photon-loss erasure.

\subsection{Coherent-state Bell measurement}
\label{sec:coherent_bell}

For coherent states with the amplitude $\beta$, $\ket{\pm\beta}$, we define the unnormalized Bell states
\begin{align}
    \ket{\Phi_{\rm coh}^{\pm}}&\propto \ket{\beta}_A\ket{\beta}_B\pm\ket{-\beta}_A\ket{-\beta}_B,\\
    \ket{\Psi_{\rm coh}^{\pm}}&\propto \ket{\beta}_A\ket{-\beta}_B\pm\ket{-\beta}_A\ket{\beta}_B.
\end{align}
A balanced beam splitter acts as
\begin{equation}
    \ket{\alpha_1}_A\ket{\alpha_2}_B\rightarrow
    \ket{(\alpha_1+\alpha_2)/\sqrt{2}}_A
    \ket{(\alpha_1-\alpha_2)/\sqrt{2}}_B.
\end{equation}
The coherent-state Bell measurement, $\mathcal{B}_\beta$, is implemented by a balaced beam splitter followed by PNRD, as shown in Fig.~\ref{fig:fusion_primitives}(c). The Bell letter is determined by the output port containing photons and the sign is determined by the photon-number parity. The only inconclusive event is vacuum detection in both outputs, so the success probability is $1-e^{-2|\beta|^2}$ \cite{Jeong2001,LeeSW2013}.

This measurement mediates the entangling operation used in Sec.~\ref{sec:resource_generation}. In particular, consider two hybrid states carrying coherent ancillas,
\begin{align}
    \ket{0_L}\ket{\beta}+\ket{1_L}\ket{-\beta},\qquad
    \ket{0_L}\ket{\beta_+}+\ket{1_L}\ket{\beta_-},
\end{align}
where $\ket{\beta_\pm}\propto\ket{\beta}\pm\ket{-\beta}$. A successful $\mathcal{B}_\beta$ on the coherent ancillas implements a logical $CZ$ edge between the H-cat qubits up to a known Pauli-frame correction \cite{LeeSW2013,Jaehak2024}.

\subsection{Cat-code Bell measurement}
\label{sec:CV_fusion}

We use the even and odd optical cat states
\begin{equation}
    \ket{C_\alpha^\pm}=N_\alpha^\pm(\ket{\alpha}\pm\ket{-\alpha}),\qquad
    N_\alpha^\pm=\frac{1}{\sqrt{2(1\pm e^{-2|\alpha|^2})}}.
\end{equation}
The even and odd cats contain only even and odd photon-number components, respectively. Since a single photon loss applies the annihilation operator and flips this parity, the loss can be detected from photon-number-resolved measurement patterns \cite{Leghtas2013,Mirrahimi2014,Jaehak2024}. 

For the H-cat encoding, the unnormalized CV Bell states are
\begin{align}
    \ket{\widetilde{\Phi}_C^\pm}&=
    \ket{C_\alpha^+}\ket{C_\alpha^+}\pm
    \ket{C_{i\alpha}^+}\ket{C_{i\alpha}^+},\\
    \ket{\widetilde{\Psi}_C^\pm}&=
    \ket{C_\alpha^+}\ket{C_{i\alpha}^+}\pm
    \ket{C_{i\alpha}^+}\ket{C_\alpha^+}.
\end{align}
We consider two different schemes for cat-code Bell measurement, which we refer to as the HA scheme~\cite{Hastrup2022} and SDR scheme~\cite{Su2022}, respectively. Both infer the CV Bell state from the photon-number pattern, denoted by $ m_C $, recorded by PNR detectors, but some patterns are compatible with more than one Bell letter. This pattern dependence is the origin of the outcome-informed error probabilities used by our decoding strategy.

\subsubsection{HA scheme}

The HA scheme uses one balanced beam splitter and two PNR detectors, as shown in Fig.~\ref{fig:fusion_primitives}(d) \cite{Hastrup2022}. Up to normalization, the no-loss Bell states transform as
\begin{widetext}
\begin{align}
\ket{\widetilde{\Phi}_C^\pm}\rightarrow&
\left(\ket{C_{\sqrt{2}\alpha}^+}_A\pm\ket{C_{i\sqrt{2}\alpha}^+}_A\right)\ket{0}_B
+\ket{0}_A\left(\ket{C_{\sqrt{2}\alpha}^+}_B\pm\ket{C_{i\sqrt{2}\alpha}^+}_B\right),\\
\ket{\widetilde{\Psi}_C^\pm}\rightarrow&
\ket{C_{i\tilde{\alpha}}^\pm}_A\ket{C_{\tilde{\alpha}}^\pm}_B
+\ket{C_{\tilde{\alpha}}^\pm}_A\ket{C_{i\tilde{\alpha}}^\pm}_B,
\end{align}
\end{widetext}
where $\tilde{\alpha}=e^{-i\pi/4}\alpha$. The photon number modulo four fixes the $XX$ fusion information. The $ZZ$ cat fusion outcome is also identified for most patterns, but the patterns $ (n_A,n_B) = (4n,0)\textrm{ or }(0,4n) $ are shared by both Bell-letter sectors. For such patterns, the post-measurement state takes the form
\begin{equation}
    \mu\ket{\Phi_C^\pm}+\nu\ket{\Psi_C^\pm}.
\end{equation}
We infer the component with the larger probability and assign the conditional fusion-outcome error
\begin{equation}
    p^{ZZ}(m_C)=\frac{|\nu|^2}{|\mu|^2+|\nu|^2}
\end{equation}
when $|\mu|\ge |\nu|$, with the labels exchanged otherwise. 
Thus Bell-letter ambiguity leads to a non-zero $ZZ$ fusion outcome error rate, whereas the phase faults generated by uncorrected multiple photon loss appear as $XX$ fusion outcome errors. This is the same physics denoted by the single-qubit $X$- and $Z$-error labels, respectively, in Ref.~\cite{Jaehak2024}.

A single photon loss changes one input parity from even to odd and is detectable from the total photon-number parity. Two or more losses can return the total parity to the no-loss sector. In particular, one loss on each CV input can either be recognized from relative photon numbers modulo four or remain indistinguishable from an even-even input, depending on the measurement pattern. The latter case produces an $XX$-fusion-outcome error. These pattern-resolved cases are included explicitly in the conditional probabilities described in Sec.~\ref{sec:outcome_model}.

\subsubsection{SDR scheme}

The SDR scheme, illustrated in Fig.~\ref{fig:fusion_primitives}(e), uses two ancillary vacuum modes, four beam splitters, one $\pi/2$ phase shifter, and four PNR detectors \cite{Su2022}. For coherent-state inputs,
\begin{align}
&\ket{\alpha_1}_A\ket{\alpha_2}_B\ket{0}_{A'}\ket{0}_{B'}\nonumber\\
&\;\rightarrow
\ket{\tfrac{\alpha_1+\alpha_2}{2}}_A
\ket{\tfrac{\alpha_1-\alpha_2}{2}}_B
\ket{\tfrac{-\tilde{\alpha}_1+i\tilde{\alpha}_2}{2}}_{A'}
\ket{\tfrac{i\tilde{\alpha}_1-\tilde{\alpha}_2}{2}}_{B'},
\end{align}
with $\tilde{\alpha}_j=e^{-i\pi/4}\alpha_j$. The $ZZ$ fusion outcome is inferred from which pair of ports contains a vacuum event, while the total detected photon number modulo four determines the $XX$ fusion result. Ambiguity appears when vacuum events occur in both the primed and unprimed sectors. As for HA, relative photon numbers modulo four also reveal some odd-odd input sectors, whereas the remaining odd-odd patterns lead to unheralded $XX$-fusion-outcome errors.

For later use, the measurement-pattern-averaged Bell-letter ambiguity probabilities conditioned on the input parities are
\begin{widetext}
\begin{align}
 p^{ZZ}_{EE,\mathrm{HA}}(\alpha)
 &=\frac{1}{2}\sech^2|\alpha|^2\left(\cosh|\alpha|^2+\cos|\alpha|^2-1\right),\\
 p^{ZZ}_{EO,\mathrm{HA}}(\alpha)
 &=\frac{1}{2}\sech|\alpha|^2,\\
 p^{ZZ}_{OO,\mathrm{HA}}(\alpha)
 &=\frac{1}{2}\csch^2|\alpha|^2\left(\cosh|\alpha|^2-\cos|\alpha|^2\right),\\
 p^{XX}_{OO,\mathrm{HA}}(\alpha)
 &=\frac{1}{2}+\frac{1}{2}\csch^2|\alpha|^2\left(\cosh|\alpha|^2-\cos|\alpha|^2\right),
\end{align}
and, for the SDR scheme,
\begin{align}
 p^{ZZ}_{EE,\mathrm{SDR}}(\alpha)
 &=\sech^2|\alpha|^2\left(\cosh\frac{3|\alpha|^2}{2}-\frac{1}{2}\cosh|\alpha|^2+\cos\frac{|\alpha|^2}{2}-\frac{1}{2}\right),\\
 p^{ZZ}_{EO,\mathrm{SDR}}(\alpha)
 &=\sech\frac{|\alpha|^2}{2}+\frac{1}{2}\sech|\alpha|^2\left(\sech^2\frac{|\alpha|^2}{2}-1\right),\\
 p^{ZZ}_{OO,\mathrm{SDR}}(\alpha)
 &=\csch^2|\alpha|^2\left(\cosh\frac{3|\alpha|^2}{2}-\frac{1}{2}\cosh|\alpha|^2-\cos\frac{|\alpha|^2}{2}+\frac{1}{2}\right),\\
 p^{XX}_{OO,\mathrm{SDR}}(\alpha)
 &=\csch^2|\alpha|^2\left(\frac{1}{2}\cosh 2|\alpha|^2-\frac{1}{2}\cosh|\alpha|^2+\cosh\frac{|\alpha|^2}{2}-\cos\frac{|\alpha|^2}{2}\right).
\end{align}
\end{widetext}
These averaged expressions are useful for the outcome-unaware decoder. The outcome-aware decoder instead retains the full PNR pattern, as described below.

\section{Error model and Decoding Strategy}

\subsection{Pure-loss evolution}
\label{sec:loss_model}

For threshold simulations we assume independent pure-loss channels with identical loss rate $\eta$ on every optical mode. Resource-state preparation is otherwise ideal, while the intrinsic fusion failure and Bell-measurement ambiguity are retained, as in the main text. The pure-loss master equation is \cite{Phoenix1990,Jaehak2024}
\begin{equation}
    \frac{d\rho}{dt}=\gamma\sum_j\left(
    \hat a_j\rho\hat a_j^\dagger
    -\frac{1}{2}\hat a_j^\dagger\hat a_j\rho
    -\frac{1}{2}\rho\hat a_j^\dagger\hat a_j\right),
\end{equation}
with $\eta=1-e^{-\gamma t}$. In the coherent-state basis,
\begin{equation}
    \ket{\alpha}\bra{\beta}\rightarrow
    \braket{\beta|\alpha}^{\eta}
    \ket{\sqrt{1-\eta}\,\alpha}\bra{\sqrt{1-\eta}\,\beta}.
\end{equation}

For an H-cat state $\ket{\psi}=a\ket{+}\ket{C_\alpha^+}+b\ket{-}\ket{C_{i\alpha}^+}$, define $\alpha'=\sqrt{1-\eta}\,\alpha$. Following the loss-sector decomposition of Ref.~\cite{Jaehak2024},
\begin{equation}
 \rho=(1-\eta)\sum_{l=0}^{3}A_l\ket{\psi'_{(l)}}\bra{\psi'_{(l)}}
 +\eta\left(B_0\tau_{(0)}+B_1\tau_{(1)}\right),
\end{equation}
where
\begin{align}
\ket{\psi'_{(0)}}&=a\ket{+}\ket{C_{\alpha'}^+}+b\ket{-}\ket{C_{i\alpha'}^+},\\
\ket{\psi'_{(1)}}&=a\ket{+}\ket{C_{\alpha'}^-}+ib\ket{-}\ket{C_{i\alpha'}^-},\\
\ket{\psi'_{(2)}}&=a\ket{+}\ket{C_{\alpha'}^+}-b\ket{-}\ket{C_{i\alpha'}^+},\\
\ket{\psi'_{(3)}}&=a\ket{+}\ket{C_{\alpha'}^-}-ib\ket{-}\ket{C_{i\alpha'}^-},
\end{align}
and $\tau_{(0)}$ and $\tau_{(1)}$ are the incoherent states obtained after loss of the DV photon with, respectively, even and odd CV parity. The coefficients are
\begin{align}
A_0&=\frac{\cosh|\alpha'|^2}{2\cosh|\alpha|^2}
\left[\cosh(\eta|\alpha|^2)+\cos(\eta|\alpha|^2)\right],\\
A_1&=\frac{\sinh|\alpha'|^2}{2\cosh|\alpha|^2}
\left[\sinh(\eta|\alpha|^2)+\sin(\eta|\alpha|^2)\right],\\
A_2&=\frac{\cosh|\alpha'|^2}{2\cosh|\alpha|^2}
\left[\cosh(\eta|\alpha|^2)-\cos(\eta|\alpha|^2)\right],\\
A_3&=\frac{\sinh|\alpha'|^2}{2\cosh|\alpha|^2}
\left[\sinh(\eta|\alpha|^2)-\sin(\eta|\alpha|^2)\right],\\
B_0&=\frac{\cosh|\alpha'|^2\cosh(\eta|\alpha|^2)}{\cosh|\alpha|^2},\\
B_1&=\frac{\sinh|\alpha'|^2\sinh(\eta|\alpha|^2)}{\cosh|\alpha|^2}.
\end{align}
Thus $B_0=A_0+A_2$ and $B_1=A_1+A_3$ are the probabilities that the CV component lies in the even- and odd-parity sectors. The four sectors $l=0,1,2,3$ reflect the fourfold cycle under photon subtraction; two additional losses return the parity but can introduce an undetected phase fault \cite{Bergmann2016, Jaehak2024}.

\subsection{Outcome-unaware and outcome-aware decoding}
\label{sec:outcome_model}
If the measurement record is discarded, the cat-fusion error probabilities are obtained by averaging over the loss sectors and PNR patterns. 
Using $B_0=A_0+A_2$ and $B_1=A_1+A_3$ with $\alpha' = \sqrt{1-\eta}\alpha$, define
\begin{equation}
    \pi_{EE}=B_0^2,\qquad \pi_{EO}=2B_0B_1,\qquad \pi_{OO}=B_1^2.
\end{equation}
For either HA or SDR, define
\begin{equation}
 p^{ZZ}_{C}(\alpha,\eta)=\pi_{EE}p^{ZZ}_{EE}(\alpha')+\pi_{EO}p^{ZZ}_{EO}(\alpha')+\pi_{OO}p^{ZZ}_{OO}(\alpha').
\end{equation}
Combining this with survival or loss of the two DV photons gives the measurement-averaged hybrid $ZZ$-fusion-outcome error
\begin{equation}
 \bar{p}^{ZZ}=\frac{1+2\eta-\eta^2}{2}\,p^{ZZ}_C(\alpha,\eta).
\end{equation}
The corresponding averaged $XX$-fusion-outcome error is obtained by summing the multiple-loss sectors,
\begin{align}
 \bar{p}_{XX}=&\frac{1}{2}\left[1-(1-\eta)^2\right]
 +(1-\eta)^2\Big[2(A_0+A_1)(A_2+A_3)\nonumber\\
 &\qquad +(A_1-A_3)^2p^{XX}_{OO}(\alpha')\Big].
\end{align}
These quantities are the uniform edge probabilities used by the outcome-unaware decoder.

Instead of replacing all outcomes by a single marginal error rate, the outcome-aware simulation retains the information that is actually available from each hybrid-fusion measurement outcomes, $m = (m_D, m_C)$, where $m_D, \, m_C$ refers to as the measurement outcome from the DV, CV fusion, respectively.  It is useful to separate the latent variables sampled in the Monte Carlo simulation from the observed variables supplied to the decoder.  For each physical hybrid fusion we first sample the DV-fusion status
\begin{equation}
    s\in\{\mathrm{S},\mathrm{F},\mathrm{L}\},
\end{equation}
corresponding to Type-II success, Type-II failure, and DV-photon loss, respectively, together with the CV input-parity sector
\begin{equation}
    q\in\{EE,EO,OO\}.
\end{equation}
The PNR record distinguishes whether the two CV parities are different, $r=\mathrm{Diff}$, corresponding to $q=EO$, or the same, $r=\mathrm{Same}$, for which the latent sector may be either $EE$ or $OO$.  Thus $EE$ and $OO$ are sampled separately in the physical error model, but are marginalized when a decoding weight is assigned to a same-parity measurement record.

The simulation samples the joint DV-status/CV-parity distribution
\begin{align}
 \mathrm{Prob}(\mathrm{S},q)&=\frac{(1-\eta)^2}{2}\,\pi_q,\\
 \mathrm{Prob}(\mathrm{F},q)&=\frac{(1-\eta)^2}{2}\,\pi_q,\\
 \mathrm{Prob}(\mathrm{L},q)&=(2\eta-\eta^2)\,\pi_q.
\end{align}
The first two lines express the $50\%$ Type-II success probability conditioned on survival of both DV photons, while the last line corresponds to loss of at least one of the two DV photons.  These probabilities sum to unity because $B_0+B_1=1$.

\subsubsection{Conditional $ZZ$-fusion-outcome probability}

For the $ZZ$ fusion, let $\mathbb{A}$ be a set of events whose measurement pattern is ambiguous to determine $ZZ$ fusion outcome, and $\mathbb{F}$ be a set of $ZZ$ fusion flipped events. Note that every $ZZ$ fusion error event happens only when measurement pattern is in $\mathbb{A}$, i.e, $\mathbb{F} \subset \mathbb{A}$.

Denoting $p^{ZZ}_{q}(\alpha')$ as a $ZZ$ fusion ambiguity probability conditioned on an input parity, $q \in \{EE, EO, OO\}$, the Monte Carlo probabilities of the three latent classes for the HA scheme and a non-successful DV fusion are
\begin{equation}
\begin{array}{c|ccc}
 & \mathbb{A} \, \cap \, \mathbb{F} \;  & \; \mathbb{A} \, \cap \, \mathbb{F}^c \; & \; \mathbb{A}^c\\ \hline
 q & \;\; p^{ZZ}_{q}\;  &  \frac12 & \;\;\;\;\frac12-p^{ZZ}_{q} 
\end{array}.
\label{eq:HA_ZZ_latent}
\end{equation}
Consequently, if the observed HA pattern is ambiguous, Bayes' rule gives
\begin{equation}
 p^{ZZ}_{\mathrm{Diff}}
 =P(ZZ\text{-flip}\mid \mathrm{ambiguous},EO)
 =\frac{p^{ZZ}_{EO}}{p^{ZZ}_{EO}+1/2}.
\label{eq:pzz_diff_HA}
\end{equation}
For a same-parity record the decoder cannot distinguish $EE$ from $OO$.  Marginalizing those two sectors with their physical priors gives
\begin{equation}
p^{ZZ}_{\mathrm{Same}}
 =\frac{\sum_{q \in \{EE,OO\}} \pi_q p^{ZZ}_q}
 {\sum_{q \in \{EE,OO\}} \pi_q (p^{ZZ}_q+1/2)}.
\label{eq:pzz_same_HA}
\end{equation}
The multiplicative factors associated with $s=\mathrm{F}$ or $s=\mathrm{L}$ cancel between the numerator and denominator, so the same posterior is used for failure and loss. 

For the SDR scheme, the corresponding latent probabilities are instead
\begin{equation}
\begin{array}{c|ccc}
 & \mathbb{A} \, \cap \, \mathbb{F} \quad & \mathbb{A} \, \cap \, \mathbb{F}^c & \quad  \mathbb{A}^c\\ \hline
 q & p^{ZZ}_q & p^{ZZ}_q & \quad 1-2p^{ZZ}_q
\end{array}.
\label{eq:SDR_ZZ_latent}
\end{equation}
Therefore an ambiguous SDR pattern has
\begin{equation}
    p^{ZZ}_{\mathrm{Same}} = p^{ZZ}_{\mathrm{Diff}}  =\frac12.
\end{equation}

Combining all information together, the hybrid fusion $ZZ$ error probability is assigned as $p^{ZZ}_{\mathrm{Same}}$ or $ p^{ZZ}_\mathrm{Diff}$ whenever $m_D \in \{F, L\}$ and $m_C \in \mathbb{A}$ according to the CV fusion input parity.
A successful Type-II fusion, or a PNR pattern classified as $m_C \notin\mathbb{A}$, gives $p^{ZZ}=0$ because the hybrid decision rule removes the Bell-letter ambiguity in those cases.

\subsubsection{Conditional $XX$-fusion-outcome probability}

The $XX$ fusion is controlled by multiple-loss sectors.  Conditioned on survival of the DV photons and on a definite latent CV parity sector, the flip probabilities implemented in the simulation are
\begin{align}
 u_{EE}&=\frac{2A_0A_2}{(A_0+A_2)^2},\\
 u_{EO}&=\frac{A_1A_2+A_0A_3}{(A_0+A_2)(A_1+A_3)},\\
 u_{OO}&=\frac{(A_1^2+A_3^2)p^{XX}_{OO}
 +2A_1A_3(1-p^{XX}_{OO})}{(A_1+A_3)^2},
\label{eq:xx_latent_parity}
\end{align}
where $p^{XX}_{OO}$ is the odd--odd cat-fusion phase-error probability for the chosen HA or SDR measurement.  The $EO$ sector is directly identified by the different-parity record, giving
\begin{equation}
    p^{XX}_{\mathrm{Diff}}=u_{EO}.
\end{equation}
For a same-parity record, the $EE$ and $OO$ sectors are again marginalized,
\begin{equation}
    p^{XX}_{\mathrm{Same}}
    =\frac{B_0^2u_{EE}+B_1^2u_{OO}}{B_0^2+B_1^2}.
\label{eq:pxx_same}
\end{equation}
These expressions are used for $s=\mathrm{S}$ and $s=\mathrm{F}$.  If at least one DV photon is lost, however, the Type-II measurement contains no reliable $XX$ information.  The simulation then labels the outcome as $XX$-undetermined, samples the actual $XX$ error as a Bernoulli variable of probability $1/2$, and assigns
\begin{equation}
    p^{XX}_{\mathrm{L}}=\frac12,
\end{equation}
which is a heralded erasure and hence has zero MWPM log-likelihood weight.

Equations~\eqref{eq:pzz_diff_HA}--\eqref{eq:pxx_same} are the conditional probabilities actually used to assign the outcome-aware decoding weights.  In particular, the true flip/nonflip variable sampled in the Monte Carlo run is not revealed to the decoder; only the experimentally accessible DV status, parity relation, and PNR-pattern class are used when computing a weight.

\subsubsection{Conversion to syndrome-graph weights}

For a single syndrome-graph edge with conditional flip probability $p_e^E$, $E\in\{XX,ZZ\}$, we use the log-likelihood-ratio weight
\begin{equation}
    w_e^E=\log\frac{1-p_e^E}{p_e^E}.
\end{equation}
A probability $p=0$ corresponds to an edge that is known to be clean and is excluded from the matching graph, whereas $p=1/2$ gives zero weight and is treated as an erasure-like edge. While the outcome-unaware decoder uses the identical marginal error probability for all edges, the outcome-aware decoder utilizes the conditional error probability corresponding to each measurement pattern.

The mapping from physical hybrid fusions to syndrome-graph edges is architecture dependent.  In the \sixring\ simulation, a physical fusion supplies a shared pair $(s,q)$ to its $ZZ$ and $XX$ fusions; conditioned on this shared pair the latent $ZZ$ and $XX$ errors are sampled independently.  The corresponding primal/dual edge weights are then obtained directly from the conditional probabilities above.

For the \fourstar\ network, one effective lattice edge is the parity of four physical sub-fusion errors.  Label the conditional error probabilities of those four sub-fusions by $p_i$.  If the four latent error bits are conditionally independent given their individual measurement records, the probability that their XOR is odd is
\begin{equation}
    p_{\mathrm{odd}}
    =\frac{1-\prod_{i=1}^{4}(1-2p_i)}{2}.
\label{eq:fourstar_odd}
\end{equation}
For a primal edge the four probabilities entering Eq.~\eqref{eq:fourstar_odd} are the $ZZ$ probabilities of sub-fusions 1 and 2 and the $XX$ probabilities of sub-fusions 3 and 4; for the dual edge the roles are reversed.  The \fourstar\ MWPM weight is then
\begin{equation}
    w=\log\frac{1-p_{\mathrm{odd}}}{p_{\mathrm{odd}}}.
\end{equation}
This architecture-specific combination is implemented before decoding and is the reason the outcome-aware \fourstar\ simulation does not simply assign one single-fusion probability to every lattice edge.

\section{Resource-state generation}
\label{sec:resource_generation}

\begin{figure*}[t]
    \centering
    \includegraphics[clip=true,width=\textwidth]{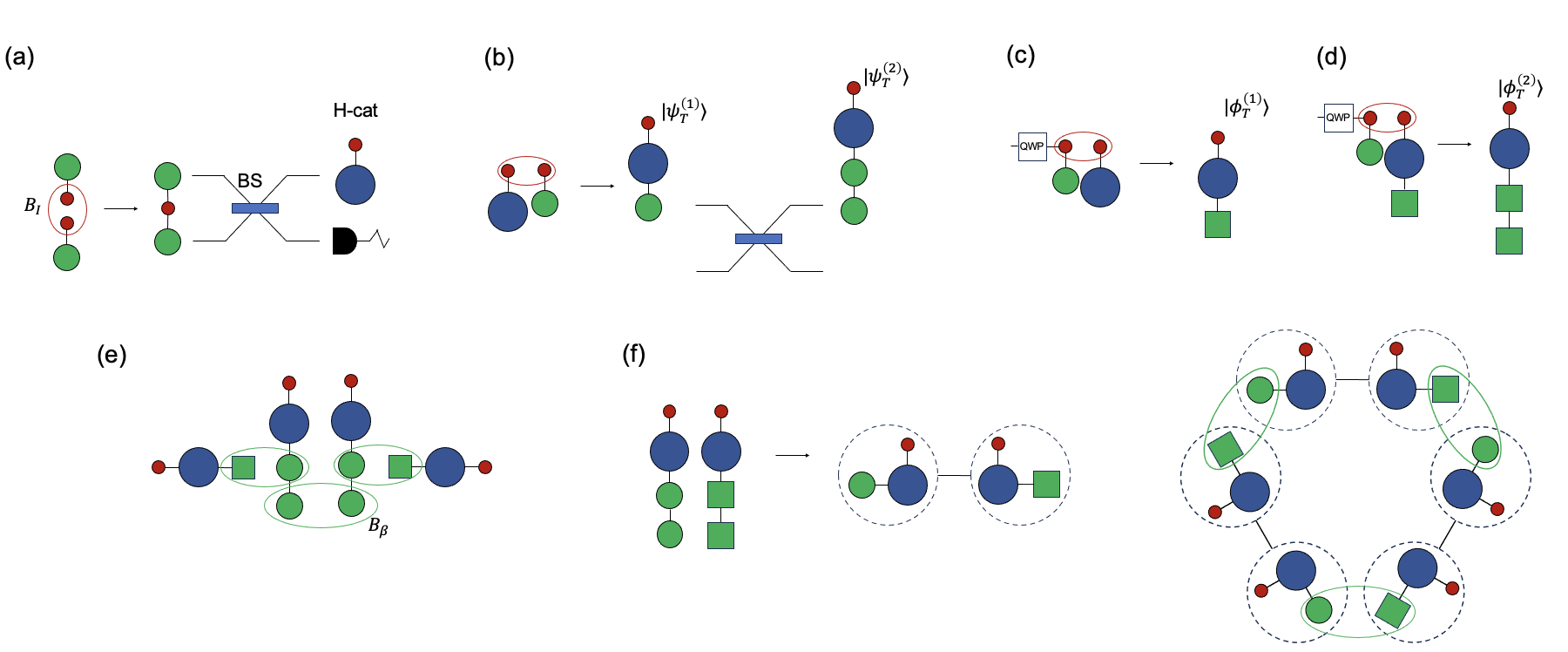}
    \caption{Detailed resource-state-generation circuits. Red, green, and blue dots denote single-photon polarization, coherent-state, and cat-code modes, respectively. The same Type-I fusion symbol is used throughout, with local polarization rotations implicit; the correponding effective Kraus operators are given in Eqs. (1) and (67). (a) H-cat generation from two phase-rotated H-coh states. (b) Generation of $\ket{\psi_T^{(1)}}$ and $\ket{\psi_T^{(2)}}$. (c) Generation of $\ket{\phi_T^{(1)}}$. (d) Generation of $\ket{\phi_T^{(2)}}$. (e) Construction of the modified \fourstar\ resource state. (f) Construction of the \sixring\ resource state.}
    \label{fig:resource_generation_sm}
\end{figure*}

We take the optical hybrid coherent-state qubit
\begin{equation}
    \ket{\mathrm{H\mbox{-}coh}}=
    \frac{\ket{+}\ket{\alpha}+\ket{-}\ket{-\alpha}}{\sqrt{2}}
\end{equation}
as the primitive resource unit \cite{LeeSW2013,kwon2015,Omkar2020,Omkar2021,Jaehak2024}. Such DV-coherent state hybrid entanglement can in principle be generated through nonlinear optical operations, including a conditional cross-Kerr interaction between a DV qubit and a coherent state~\cite{Jeong2005,Nemoto2004}, or through photon addition operations~\cite{Jeong2014}. Alternatively, it can be generated in a measurement-induced manner using non-Gaussian states, linear-optical operations, and conditional photodetection~\cite{kwon2015,Morin2014}. Experimental demonstrations of related hybrid entanglement generation and processing have been reported using linear-optical and measurement-induced techniques~\cite{Morin2014,Guccione2020,Darras2023}.

The only nontrivial probabilistic operations used below are the Type-I fusion $\mathcal{B}_I$, with success probability $1/2$, and the coherent-state Bell measurement $\mathcal{B}_\beta$, whose success probability is
\begin{equation}
    P_{\mathcal{B}_\beta}=1-e^{-2|\beta|^2}.
\end{equation}
A successful $\mathcal{B}_\beta$ between a coherent-basis ancilla $\{|\beta\rangle,|-\beta\rangle\}$ and a cat-basis ancilla $\{|\beta_+\rangle,|\beta_-\rangle\}$ implements a logical $CZ$ edge up to a known Pauli-frame correction.  For example, suppressing normalization,
\begin{align}
 \ket{\chi_1}&=\ket{0}\ket{\beta}+\ket{1}\ket{-\beta},\\
 \ket{\chi_2}&=\ket{0}(\ket{\beta}+\ket{-\beta})
 +\ket{1}(\ket{\beta}-\ket{-\beta}),
\end{align}
are projected by a successful coherent Bell measurement onto a state locally equivalent to
\begin{equation}
    \ket{00}+\ket{01}+\ket{10}-\ket{11},
\end{equation}
i.e., $CZ\ket{+}\ket{+}$.  This motivates attaching coherent ancilla modes to the logical H-cat qubits before assembling the FBQC graph state. 

To generate the resource state for our protocol, we first devised a scheme to build an H-cat state in the logical $X$ basis. Then we attach the $n$-number of ancillary coherent states to the H-cat. According to the basis of ancillary states, we refer to as these intermediate states as $|\psi_T^{(n)}\rangle$ and $ |\phi_T^{(n)}\rangle$, corresponding to a coherent- and a cat-basis, respectively. Finally, consuming coherent ancillary states by $\mathcal{B}_\beta$, we apply logical $CZ$ gate, generating the desired resource state. The full sequence is summarized in Fig.~\ref{fig:resource_generation_sm}.

\subsection{H-cat preparation}

To generate an H-cat state, two H-coh states with phase-rotated coherent amplitudes $e^{i\pi/4}\alpha$ and $e^{-i\pi/4}\alpha$ are first fused by $\mathcal{B}_I$, leaving one DV qubit. Their coherent components are then mixed on a balanced beam splitter and one output is measured by a PNR detector.  The beam splitter creates the required two-component coherent superpositions; conditioning on an even detected photon number produces a state in the H-cat logical subspace.  Outcomes $0\pmod 4$ and $2\pmod 4$ differ only by a known logical Pauli correction, which can be absorbed into the DV wave-plate/Pauli frame.  Because two H-coh states are consumed, $\mathcal{B}_I$ succeeds with probability $1/2$, and the useful even-parity postselection is $(1+ e^{-2|\alpha|^2})/2$ which approaches probability $1/2$ for the amplitudes of interest, the mean H-cat cost is approximately
\begin{equation}
    N_{\mathrm{H\mbox{-}cat}}\simeq 2\times2\times2=8
\end{equation}
H-coh states.

\subsection{Generation of $\ket{\psi_T^{(1)}}$ and $\ket{\psi_T^{(2)}}$}

To mediate graph-state entangling operations, we first prepare a logical H-cat qubit with one coherent ancilla,
\begin{equation}
 \ket{\psi_T^{(1)}(\beta)}
 =\ket{0_L}\ket{\beta}+\ket{1_L}\ket{-\beta}.
\label{eq:psiT1}
\end{equation}
Operationally, the H-cat polarization mode is put in the $H/V$ basis appropriate for the Type-I projector and fused with the polarization mode of an H-coh state. With Hadamard rotations applied to both input polarization modes, the Type-I fusion defined in Eq. (1) has, up to the known output-dependent wave-plate correction, the effective successful Kraus operator
\begin{equation}
    K_I=\ket{+}\!\bra{++}+\ket{-}\!\bra{--}
\end{equation}
which selects the equal-polarization branches and produces Eq.~\eqref{eq:psiT1}.  Since the inputs cost $8+1$ H-coh states and the Type-I fusion succeeds with probability $1/2$,
\begin{equation}
    N_{\psi_T^{(1)}}=2(8+1)=18.
\end{equation}

The second-order state is generated without another probabilistic fusion.  We first prepare Eq.~\eqref{eq:psiT1} with coherent amplitude $\sqrt{2}\beta$,
\begin{equation}
 \ket{\psi_T^{(1)}(\sqrt{2}\beta)}
 =\ket{0_L}\ket{\sqrt{2}\beta}
 +\ket{1_L}\ket{-\sqrt{2}\beta},
\end{equation}
and mix its coherent ancilla with vacuum on a balanced beam splitter.  Using
\begin{equation}
    \ket{\pm\sqrt{2}\beta}\ket{0}
    \longrightarrow
    \ket{\pm\beta}\ket{\pm\beta},
\end{equation}
we obtain
\begin{equation}
 \ket{\psi_T^{(2)}}
 =\ket{0_L}\ket{\beta}\ket{\beta}
 +\ket{1_L}\ket{-\beta}\ket{-\beta}.
\label{eq:psiT2}
\end{equation}
Thus the balanced beam splitter increases the number of coherent ancilla modes from one to two without consuming an additional H-coh resource, and
\begin{equation}
    N_{\psi_T^{(2)}}=N_{\psi_T^{(1)}}=18.
\end{equation}
More generally, the superscript in $\ket{\psi_T^{(n)}}$ denotes the number of coherent ancilla modes attached to the logical H-cat qubit.

\subsection{Generation of $\ket{\phi_T^{(1)}}$ and $\ket{\phi_T^{(2)}}$}

A second ancillary family is required because $\mathcal{B}_\beta$ connects a coherent-sign basis to the even/odd coherent-cat basis.  We define
\begin{equation}
 \ket{\phi_T^{(n)}}
 =\ket{0_L}\ket{\beta_+}^{\otimes n}
 +\ket{1_L}\ket{\beta_-}^{\otimes n},
 \quad
 \ket{\beta_\pm}\propto\ket{\beta}\pm\ket{-\beta}.
\end{equation}
The first-order state is obtained by applying a quarter wave-plate rotation to the DV mode of an H-coh state followed by $\mathcal{B}_I$ with an H-cat state.  In the resulting basis,
\begin{equation}
 \ket{\phi_T^{(1)}}
 =\ket{0_L}\ket{\beta_+}
 +\ket{1_L}\ket{\beta_-},
\end{equation}
so no additional probabilistic optical operation is required and its mean cost remains 18 H-coh states. Unlike $|\psi_T^{(2)}\rangle$, a sequential Type-I measurement between polaraization rotated H-coh and $|\phi_T^{(1)}\rangle$ yields the second-order state, $|\phi_T^{(2)}\rangle$.  Its mean resource cost is therefore
\begin{equation}
    N_{\phi_T^{(2)}}=2\left(N_{\psi_T^{(1)}}+1\right)
    =2(18+1)=38
\end{equation}
H-coh states.

\subsection{Assembly of the \fourstar\ and \sixring\ resources}

Once the $\psi_T$ and $\phi_T$ building blocks are available, the coherent ancillae are consumed pairwise by successful $\mathcal{B}_\beta$ measurements.  Each such measurement adds a logical $CZ$ edge (up to a Pauli-frame correction) while removing the measured coherent ancilla modes.

The modified \fourstar\ resource used in this work has stabilizer generators
\begin{equation}
    \{X_1X_2Z_3Z_4,\;Z_1Z_2,\;Z_2X_3,\;X_3X_4\},
\end{equation}
with fusion basis $\{XX,ZZ\}$.  It is assembled from two $\ket{\psi_T^{(2)}}$ states and two $\ket{\phi_T^{(1)}}$ states.  Four coherent Bell measurements are required, so the mean H-coh cost is
\begin{align}
 N_{\fourstar}
 &=\left(2\times18+2\times18\right)
 \left(1-e^{-2|\beta|^2}\right)^{-4}\\
 &=72\left(1-e^{-2|\beta|^2}\right)^{-4}.
\end{align}

The \sixring\ is assembled from three $\ket{\psi_T^{(2)}}$ states and three $\ket{\phi_T^{(2)}}$ states arranged cyclically.  The six adjacent coherent-ancilla pairs are fused by six $\mathcal{B}_\beta$ measurements, producing the six logical graph edges of the ring.  Its mean H-coh cost is therefore
\begin{align}
 N_{\sixring}
 &=\left(3\times18+3\times38\right)
 \left(1-e^{-2|\beta|^2}\right)^{-6}\\
 &=168\left(1-e^{-2|\beta|^2}\right)^{-6}.
\end{align}
\putbib
\end{bibunit}

\end{document}